Journal of Insurance and Financial Management, Vol. 1, Issue 5 (2016) 92-123

JIFM  JOURNAL OF INSURANCE AND FINANCIAL MANAGEMENT

# Artificial Neural Network and Time Series Modeling Based Approach to Forecasting the Exchange Rate in a Multivariate Framework

Tamal Datta Chaudhuri [a], Indranil Ghosh [b,*]

[a,b] Calcutta Business School, Diamond Harbour Road, Bishnupur – 743503, 24 Paraganas (South), West Bengal, India

ABSTRACT

Any discussion on exchange rate movements and forecasting should include explanatory variables from both the current account and the capital account of the balance of payments. In this paper, we include such factors to forecast the value of the Indian rupee vis a vis the US Dollar. Further, factors reflecting political instability and lack of mechanism for enforcement of contracts that can affect both direct foreign investment and also portfolio investment, have been incorporated. The explanatory variables chosen are the 3 month Rupee Dollar futures exchange rate (FX4), NIFTY returns (NIFTYR), Dow Jones Industrial Average returns (DJIAR), Hang Seng returns (HSR), DAX returns (DR), crude oil price (COP), CBOE VIX (CV) and India VIX (IV). To forecast the exchange rate, we have used two different classes of frameworks namely, Artificial Neural Network (ANN) based models and Time Series Econometric models. Multilayer Feed Forward Neural Network (MLFFNN) and Nonlinear Autoregressive models with Exogenous Input (NARX) Neural Network are the approaches that we have used as ANN models. Generalized Autoregressive Conditional Heteroskedastic (GARCH) and Exponential Generalized Autoregressive Conditional Heteroskedastic (EGARCH) techniques are the ones that we have used as Time Series Econometric methods. Within our framework, our results indicate that, although the two different approaches are quite efficient in forecasting the exchange rate, MLFNN and NARX are the most efficient.

Journal of Insurance and Financial Management

*Corresponding Author:
indranilg@calcuttabusinessschool.org

Journal of Insurance and Financial Management (ISSN-Canada: 2371-2112)

## 1. Introduction

Exchange rate is the price of foreign currency in terms of the domestic currency. If 1 US Dollar can buy 65 Indian Rupees, then the exchange rate is either Rs.65/$1 or $1/Rs.65. As exchange rate is a price, any analysis of it has to be based on factors affecting demand and supply of foreign currency. Foreign currency flows in and out of an economy through both the current account and the capital account of the balance of payments [for a lucid explanation see Caves and Jones (1973)]. Export and import of goods and services leads to inflow and outflow respectively of foreign currency through the current account. Portfolio flows and foreign direct investment lead to inflow or outflow through the capital account. Since balance of payments have to balance, any discrepancy is taken care of by change in the level of foreign currency reserves.

In today's world, most economies have shifted to a flexible exchange rate system where the exchange rate adjusts to clear the market. In case of abnormal movements in the rate, central banks do intervene to stabilize the currency by selling from reserves or buying from the market. Since the exchange rate is not entirely in the hands of the domestic economy, and since not all foreign currency contracts are spot contracts, exporters and importers face the risk of adverse movements in the exchange rate in the future. For this, there is a strong forward and futures market in the foreign currency market. All the above characteristics of this market has led to the emergence of three sets of players namely, agents with underlying interest in foreign currency like exporters, importers and traders, speculators and arbitrageurs.

Exports and imports of an economy depend on the economic background and the various policies for economic development, the technological capabilities, natural endowment of resources, and the sociological and demographic features. India does not have large reserves of crude oil. Hence around 80% of its imports is crude oil. Venezuela on the other hand has significant reserves of crude oil and hence it exports mostly oil. Brazil is endowed with coffee plantations and have nurtured them to make it the largest coffee producer and exporter in the world. So movement in foreign currency through the current account are dependent on factors that affect exports and imports. Movement of foreign exchange through the capital account has to do with returns on foreign exchange, and hence on the relative attractiveness of two nations in terms of returns.

Let there be two countries USA (u) and India (i), the rupee dollar spot rate be $S_p$, the rupee dollar forward rate be $S_f$ and the nominal rates of interest in the two countries be $R_u$ and $R_i$ respectively. Then, in equilibrium, the following equality has to hold.

$(1 + R_i) = S_f(1 + R_a)/S_p$

Otherwise funds will move between countries. This condition is known as the "covered interest arbitrage" condition. The condition states that, Rs.1 in India, converted at the spot rate to US dollars, invested in the US for a period and brought back to India by the period forward rate today, has to be equal to the returns from the Indian market for the period. If the LHS is greater than the RHS, then foreign funds will move into India and if RHS is greater than LHS then funds will move from India to the US. If we replace nominal rates of interest with real rates of interest and as real rates will equalize across countries, the funds movement will be governed by the rates of inflation in the two countries.

We mentioned that the returns from the two countries have to be equal in equilibrium, but this equality has to be qualified by the ratio of spot rate to the forward rate. This is the characteristic of the exchange rate market which attracts arbitrageurs. Notice that the above equality condition is one equation in four unknowns. Thus given the spot rate and the rates of return in the two countries, the forward rate can be derived. On the other hand, if we use the quoted futures rate, then one of the interest rates can be derived. This derived interest rate is called the Implied Repo Rate, and if it is different from the actual interest rate, will give rise to arbitrage.

The above discussion was intended to establish that any study on exchange rate movements and forecasting, has to include explanatory variables from both the current account and the capital account. In this paper, we include such factors to forecast the value of the Indian rupee vis a vis the US Dollar. There could be certain other factors like political instability and lack of mechanism for enforcement of contracts that can affect both direct foreign investment and also portfolio investment. In this paper we include such variables also.

To forecast the exchange rate, we have used two different classes of frameworks namely, Artificial Neural Network (ANN) based models and Time Series Econometric models. Multilayer Feed Forward Neural Network (MLFFNN) and Nonlinear Autoregressive models with Exogenous Input (NARX) Neural Network are the approaches that we have used as ANN models. Generalized Autoregressive Conditional Heteroskedastic (GARCH) and Exponential Generalized Autoregressive Conditional Heteroskedastic (EGARCH) techniques are the ones that we use as Time Series Econometric methods.

Although the existing literature on forecasting the exchange rate is quite rich and this paper is embedded in that literature, the following features of the paper will add to the understanding of the problem at hand and also refine the forecasting of future exchange rate movements. First, we use daily data of the exchange rate and the other explanatory variables. We do not incorporate any macroeconomic variables. Our contention is that the exchange rate is a price of a financial asset which is traded on real time basis. Hence its forecast should use such variables, the data on which are also available daily. Second, we use both current account and capital account factors as explanatory variables. In particular, we use the forward rate as an explanatory variable. Third, as mentioned above, we incorporate explanatory variables that reflect the economic/political/financial instability of an economy. Fourth, in contrast to many papers that have used machine learning techniques and econometric techniques in a univariate framework, ours is a multivariate approach. Fifth, we apply both traditional econometric and machine learning tools in a multivariate framework, which enables us to compare the efficiency of these two classes of models. Sixth, application of the NARX model is quite unique.

The plan of the paper is as follows. A brief literature survey is presented in Section 2. Detailed description of the dataset and methodologies are elucidated in Section 3. Results obtained from ANN modelling and time series modelling are explained in Sections 4 and 5 respectively. Comparative analysis of the performance of the two different frameworks is presented in Section 6. Section 7 concludes the paper.

## 2. Literature Review

Meese and Rogoff (1983) presented a model of forecasting the exchange rate which incorporated the characteristics of the flexible price monetary model (Frenkel-Bilson), the sticky price monetary model (Dornbusch-Frankel), and the sticky price asset model (Hooper – Morton). The relationship they postulated was

$$S = A + A_1 (M – M^*) + A_2 (Y - Y^*) + A_3 (R_s – R_s^*) + A_4 ( P – P^*) + A_5 TB + A_6 TB^* + u$$

where S is log dollar price of foreign currency, $M - M^*$ is log ratio of domestic and foreign money supply, $Y – Y^*$ is log ratio of domestic and foreign income, $R_s – R_s^*$ and $P – P^*$ are the short term interest rate and inflation rate differential respectively, TB and $TB^*$ are the foreign exchange reserves and u is the random disturbance term. Their specification is macroeconomic in nature where money supply, national income and forex reserves were considered.

Zhang and Berardi (2001) used neural network ensembles for predicting the exchange rate between the British pound and US dollar. In his paper, the inputs are the lag variables of the output and the focus is on the technique of prediction.

Perwej and Perwej (2012) considered forecasting the Indian Rs/UD$ exchange rate using the Artificial Neural Network framework. Here also the focus is on selection of number of input nodes and hidden layers. Lagged values of the exchange rate is used to predict the future values of the exchange rate.

In the lines of Meese and Rogoff (1983), Lam, Fung and Yu (2008) consider a Bayesian model for forecasting the exchange rate. The explanatory variables they consider are quite exhaustive and include stock price, change in stock price, long-term interest rate, short-term interest rate, term spread, oil price, change in oil price, exchange rate return of the previous period, sign of exchange rate return of the previous period, seasonally adjusted real GDP, change in seasonally adjusted real GDP, seasonally adjusted money supply, change in seasonally adjusted money supply, consumer price level, inflation rate, and ratio of current account to GDP. Many of the variables are measured relative to that of the foreign country. However, the forward rate is missing as an explanatory variable.

Ravi, Lal and Raj Kiran (2012) use six nonlinear ensemble architectures for forecasting exchange rates. Although the paper applies many techniques for forecasting a number of exchange rates, the input variables are only the lagged values of the exchange rate itself.

Dua and Ranjan (2011) applied both Vector Auto Regression (VAR) and Bayesian Vector Auto Regression (BVAR) for forecasting the Indian Rupee/US Dollar exchange rate. Following Meese and Rogoff (1983), they also consider both current account and capital account variables along with forward exchange rates. While they find that the BVAR outperforms VAR, they also observe that forecast accuracy improves if we include the forward premium and volatility of capital flows.

Pacelli (2012) analyzed and compared the predictive ability of ANN, ARCH and GARCH models where the output variable was the daily exchange rate Euro/Dollar and the input variables were the Nasdaq Index, Daily Exchange Rate Eur/Usd New Zealand, Gold Spot Price USA, Average returns of Government Bonds - 5 years in the USA zone, Average returns of Government Bonds - 5 years in the Eurozone, Crude Oil Price , Exchange rate Euro / US dollar of the previous day compared to the day of the output.

Imam et al. (2015) presented a comprehensive survey work on various computational intelligent methods and financial quantitative models used in predictive modelling of exchange rates. The study highlights the usage of Artificial Neural Networks, Support Vector Machine, ARCH/GARCH models etc. in the particular area.

Androu and Zombanakis (2006) utilized Artificial Neural Network based approaches to forecast the Euro exchange rate versus the United States Dollar and the Japanese Yen. Their work suggested the presence of random behavior of time series according to Rescaled Range Statistic (R/S). However the NN model adopted in their study indeed resulted in good prediction in terms of Normalized Root Mean Squared Error (NRMSE), the Correlation Coefficient (CC), the Mean Relative Error (MRE), the Mean Absolute Error (MAE) and the Mean Square Error (MSE).

Vojinovic and Kecman (2001) employed Radial Basis Function Neural Network to predict daily $US/$NZ closing exchange rates. Findings indicated that Radial Basis Function Neural Network outperformed traditional autoregressive models.

Garg (2012) presented a novel framework comprising of GARCH extended machine learning models namely, Regression trees, Random Forests, Support Vector Regression (SVR), Least Absolute Shrinkage and Selection Operator (LASSO) and Bayesian Additive Regression trees (BART) to predict EUR/SEK, EUR/USD and USD/SEK exchange rates on both monthly and daily basis in a multivariate framework where different sets of predictors were utilized for prediction of respective exchange rates.

Jena et al. (2015) used Knowledge Guided Artificial Neural Network (KGANN) structure for exchange rate prediction. Premanode and Toumazou (2013) proposed a novel methodology where differential Empirical Mode Decomposition (EMD) is utilized to improve the performance of Support Vector Regression (SVR) to forecast exchange rates. Findings suggested that the proposed framework outperformed Markov Switching GARCH and Markov Switching Regression Models. Majhi et al. (2012) made a comparative study using Wilcoxon Artificial Neural Network (WANN) and Wilcoxon Functional Link Artificial Neural Network (WFLANN) in predictive modelling of exchange rate.

There is an extant literature in finance where both machine learning tools and econometric methods have been applied. For example, Datta Chaudhuri and Ghosh (2015) applied multilayer feedforward neural network and cascaded feedforward neural network to predict volatility measured in terms of volatility in NIFTY returns and volatility in gold returns over the years. Predictors considered in the study were India VIX, CBOE VIX, volatility of crude oil returns, volatility of DJIA returns, volatility of DAX returns, volatility of Hang Seng returns and volatility of Nikkei returns. Results justified the usage of ANN based methodology.

Bhat and Nain (2014) used GARCH, EGARCH and Component GARCH (CGARCH) to critically analyze volatility measures of four different Bombay Stock Exchange (BSE) indices during January, 1, 2002 to December, 31, 2013. Findings reported that BSE IT, BSE PSU, BSE Metal and BSE Bankex exhibit clear volatility clustering. Tripathy and Rahman (2013) used GARCH (1, 1) model to measure the conditional market volatility of Bombay Stock Exchange (BSE) and Shanghai Stock Exchange (SSE). Empirical results obtained from daily data of 23 years of their study strongly indicate presence of volatility in the market.

## 3. Data and Methodology

The data set chosen for the study is daily data from 1.1.2009 to 8.4.2016 with 1783 observations. The dependent variable is the Rupee Dollar exchange rate (FX1). The independent variables are the 3 month Rupee Dollar futures exchange rate (FX4), NIFTY returns (NIFTYR), Dow Jones Industrial Average returns (DJIAR), Hang Seng returns (HSR), DAX returns (DR), crude oil price (COP), CBOE VIX (CV) and India VIX (IV). Inclusion of FX4, NIFTYR, DJIAR, HSR and DR as explanatory variables follows from the "covered interest arbitrage condition". While NIFTYR represents daily returns from the Indian stock market, we have included DJIAR and DR to represent returns from the western part of the world and HSR to represent returns in the eastern part of the world. CV and IV have been included to control for relative uncertainty in the Indian market vis a vis the US market as they are measures of Implied Volatility and are forward looking measures. As crude oil is the single largest import item of India, COP is the included to represent the current account in the balance of payments. In this study, thus, we have both current account and capital account variables, along with measures of volatility. Descriptive Statistics of all these variables are presented in the following table.

**Table 1**
Descriptive Statistics of Variables

| Variable | Mean | Median | Maximum | Minimum | Std. Dev. | Skewness | Kurtosis |
|---|---|---|---|---|---|---|---|
| FX1 | 54.6283 | 54.2300 | 69.0625 | 43.9150 | 7.5670 | 0.1598 | 1.5823 |
| FX4 | 55.0946 | 55.0588 | 69.8575 | 44.3800 | 8.0879 | 0.1485 | 1.4841 |
| NIFTYR | 0.00059 | 0.00047 | 0.17744 | -0.01681 | 0.0128 | 1.3702 | 24.3389 |
| DJIAR | 0.00043 | 0.00053 | 0.06835 | -0.05546 | 0.01037 | -0.1089 | 7.0345 |
| HSR | 0.00029 | 0.00002 | 0.05756 | -0.06461 | 0.01364 | -0.11242 | 5.2778 |
| DR | 0.00051 | 0.00089 | 0.08152 | -0.05819 | 0.01417 | -0.00099 | 5.19045 |
| COP | 0.00039 | 0.00027 | 0.26874 | -0.09007 | 0.02187 | 1.69542 | 19.9603 |
| CV | 20.2721 | 17.5950 | 137.150 | 10.3200 | 8.43918 | 2.79998 | 24.8811 |
| IV | 22.1800 | 19.9450 | 63.5800 | 11.5650 | 8.08505 | 1.71638 | 6.28577 |

The values of skewness and kurtosis measures of majority of the variables indicate the presence of leptokurtosis. Jarque-Bera test has been conducted for statistical significance. The results shown in Table 2 dispel the normality assumption at 0.1% level for all the variables, indicating the presence of volatility, thus justifying the use of GARCH and EGARCH models.

**Table 2**
Jarque-Bera Test

| Variable | Jarque-Bera | p-value | Significance |
|---|---|---|---|
| **FX1** | 156.8169 | 0.000000 | *** |
| **FX4** | 177.1663 | 0.000000 | *** |
| **NIFTYR** | 34367.26 | 0.000000 | *** |
| **DJIAR** | 1212.111 | 0.000000 | *** |
| **HSR** | 388.9902 | 0.000000 | *** |
| **DR** | 356.2574 | 0.000000 | *** |
| **COP** | 22211.97 | 0.000000 | *** |
| **CV** | 37877.93 | 0.000000 | *** |
| **IV** | 1676.577 | 0.000000 | *** |

*** Significant at 1% level

**3.1 Multilayer Feed-Forward Network (MLFFNN):** It is a standard Artificial Neural Network (ANN) technique that attempts to mimic the working nature of the human brain to extract the hidden pattern between a set of inputs and outputs (Haykin, 1999). Human brain is composed of around $10^{10}$ number of highly interconnected units known as neurons. Similarly neurons are structured and connected in a hierarchical manner in a layered architecture of ANN. There are three distinct interconnected layers in a typical ANN architecture namely, an input layer, hidden layer(s) and an output layer. They are connected via neurons and strength of each connection is actually represented by numeric weight value. For prediction tasks, basically these weight values corresponding to decision boundary, are estimated using various optimization algorithms on training data set. Once the estimated values are stabilized after validation, trained ANN is tested against a test data set to assess its predictive power.

A Multilayer Feed-Forward Neural Network (MLFFNN) is also composed of an input layer, one or more hidden layers and an output layer. A typical MLFFNN having 5 inputs is depicted below.

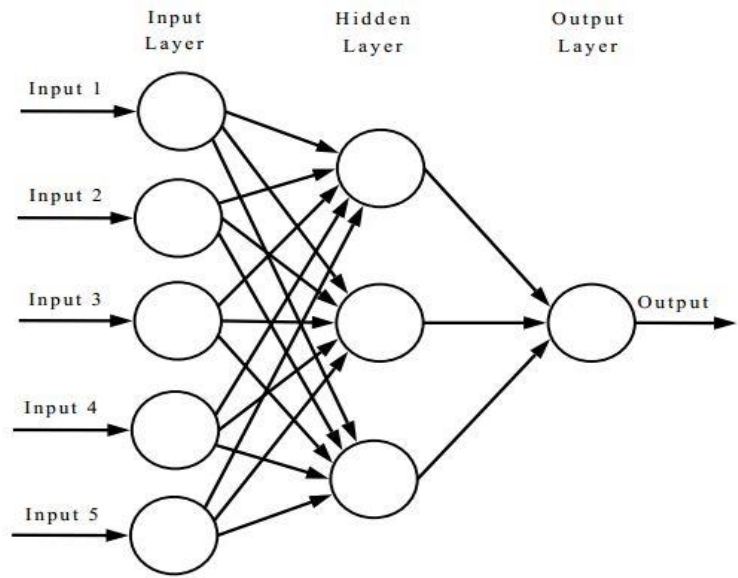

**Figure 1**
Simple architecture of MLFFNN

Input layer consists of simultaneously fed input units. Subsequently weighted inputs are fed into hidden layer. The weighted outputs of hidden layer(s) serve as the inputs to output layer and represent the prediction of network. As none of the weights cycle back to an input unit or to a previous layer's output unit and there are at least three distinct layers, this network topology is called Multilayer Feed-Forward network.

Each output unit in a particular layer of MLFFNN considers weighted sum of outputs from previous layer's as inputs. It then applies a nonlinear function to the received input and forwards them to the subsequent layer. Each input signal ($x_i$) is associated with a weight ($w_i$). The overall input I to the processing unit is a function of all weighted inputs given by

$I = f(\sum x_i \times w_i)$     (1)

The activation state of the processing unit (A) at any time is a function (usually nonlinear) of I.

$A = g(I)$     (2)

In general, logistic or sigmoid function is used as activation function. If net input to unit j is $I_j$ then output ($O_j$) of unit j may be calculated as:

$O_j = \frac{1}{1+e^{-I_j}}$ ………………………… (3)

The output Y from the processing unit is determined by the transfer function h

$Y = h(A) = h(g(I)) = h(g(f(\sum x_i \times w_i)) = \Theta(\sum x_i \times w_i)$     (4)

Given large sample of training data, MLFFNN can estimate the weight values and perform nonlinear regression. Once the estimated values are stabilized after validation, trained MLFFN can be utilized for prediction on test data set. Backpropagation algorithm has been used as training algorithm of MLFFNN for estimation of parameters to capture the nonlinear pattern between set of outputs and inputs for predictive modelling.

We now elucidate the working principle of Backpropagation Algorithm as reported by Han et al. (2009). Steps of general purpose backpropagation algorithm are outlined below.

1. Randomly initialize the weight and bias values of the network.

2. While terminating condition is not met {

3. For each training sample in dataset {

4. For each input layer unit j {

5. Output of an input unit is   $O_j = I_j$; //

6. For each hidden or output layer unit j {

7. Net input of unit j is computed as $I_j = \sum_i w_{ij} O_i + \theta_j$; // $\theta_j$ is the respective bias value

8. Output of each unit j $O_j = \frac{1}{1+e^{-I_j}}$; }

9. For each unit j in the output layer {

10. Error is calculated as: $Err_j = O_j(1-O_j)(T_j-O_j)$; // $T_j$ is the target at $j^{th}$ unit }

11. For each unit j in the hidden layer(s), from last to first hidden layer {

12. Error is computed as:  $Err_j = O_j(1 - O_j) \sum_k Err_k w_{jk}$ ; }

13. For each weight $w_{ij}$ in network {

14. Weight value as are updated as: $w_{ij} = w_{ij} + \Delta w_{ij}$ ;

15. Where $\Delta w_{ij} = (l)Err_j O_i$ ; // l denotes the learning rate }

16. For each bias $\theta_j$ in network {

17. Bias values are modified as: $\theta_j = \theta_j + \Delta \theta_j$;

18. Where $\Delta \theta_j = (l)Err_j$ ; }

19. }}

**3.2 NARX (Nonlinear Autoregressive models with exogenous input) Neural Network Model:** NARX neural network is a variant of Recurrent Network (Lin et al. 1996, Gao and Meng, 2005) that has been successfully utilized in time series prediction problems. The major difference between RNN and MLFFN is that RNN allows a weighted feedback connection between layers of neurons and thereby making it suitable for time series analysis by allowing lagged values of variables to be considered in model. Although throughout the literature many time series methods such Autoregressive (AR), Moving Average (MA), Autoregressive Moving Average (ARMA), Autoregressive Integrated Moving Average (ARIMA), etc. have been applied in various econometrics problems, these techniques cannot cope with nonlinear problems. NARX on the contrary can efficiently be used for modelling non stationary and nonlinear time series. Mathematically input output representation of nonlinear discrete time series in NARX network is governed by the following equation.

$$y(t) = f[u(t - D_u), \ldots\ldots, u(t - 1), u(t), y(t - D_y), \ldots\ldots, y(t - 1)] \ldots\ldots (5)$$

where u(t) and g(t) represent input and output of the network at time t. $D_u$ and $D_y$, are the input and output order, and the function f is a nonlinear function. The function is approximated by MLFFN. It is also possible to have NARX networks with zero input order and a one-dimensional output. i.e., having feedback from output only. In such cases operation of NARX network is governed by equation 6.

$$y(t) = \Psi[u(t), y(t - 1), \ldots\ldots y(t - D)] \quad (6)$$

where Ψ is the mapping function, approximated by standard MLFFNN. Schematic structure of NARX is depicted in figure 2.

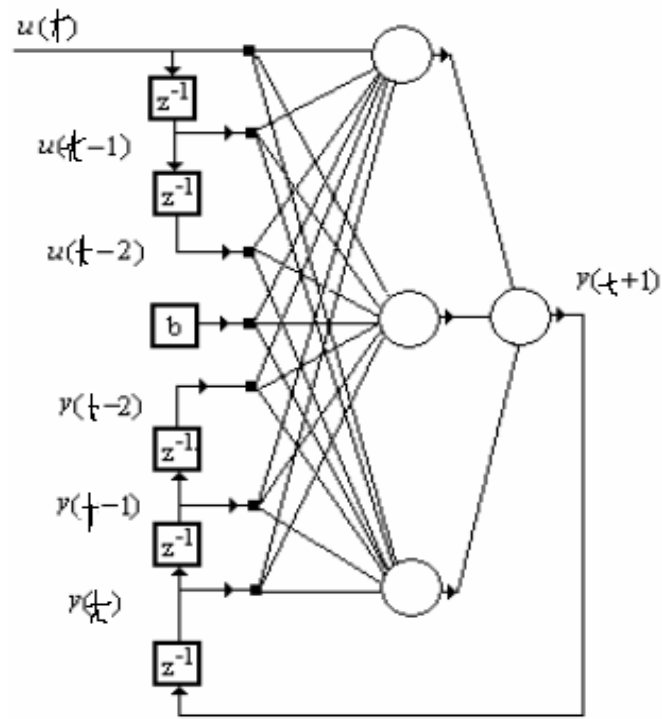

**Figure 2**
NARX architecture

**3.3 Time Series Modelling:** Objective of time series forecasting is to predict future values of a time series, $X_{n+m}$ based on the observed data to present, $X = \{X_n, X_{n-1}, ....., X_1\}$. Majority of the time series forecasting model assume $X_t$ to be stationary. Autoregressive (AR), Moving Average (MA), Autoregressive Moving Average (ARMA), Autoregressive Integrated Moving Average (ARIMA), Autoregressive Distributed Lag (ARDL), etc. are various time series modelling techniques that are predominantly applied to forecast linear and univariate time series. Nonstationary time series can be converted to stationary by various means to fit these models. However it has been observed that financial data exhibits volatility clustering i.e., high volatile periods are followed by high volatile periods and low volatile periods by low volatile periods. One of the major limitation of the above mentioned models is that they are all built under the assumption that conditional variance of past is constant. For heteroskedastic situations traditional linear time series forecasting techniques fail to capture the volatility and thereby yield poor predictions. Autoregressive Conditional Heteroskedastic (ARCH) proposed by Engle (1982) was introduced to model volatility. It has been further extended to Generalized Autoregressive Conditional Heteroskedastic (GARCH) model by Bollerslev (1986), Exponential GARCH (EGARCH) by Nelson (1990), Threshold ARCH (TARCH) by Rabemananjara and Zakoian in 1993, Quadratic ARCH by Sentana in 1995, etc.

In this study, we have applied multivariate GARCH and EGARCH model. Generalized description of both these two models are furnished below.

**3.3.1 GARCH:** GARCH (p, q) model is actually same as ARCH model of infinite order. For GARCH (p, q) model, the conditional volatility ($h_t$) is a function of previous conditional volatility ($h_{t-p}$) and previous squared error ($\varepsilon^2_{t-q}$). The standard GARCH (1, 1) for stock returns model can be represented by following equations.

$$R = c + \rho R_{t-1} + \varepsilon_t \quad \ldots\ldots\ldots\ldots\ldots\ldots\ldots\ldots(7)$$

$$\varepsilon_t = z_t \sqrt{h_t} \quad \ldots\ldots\ldots\ldots\ldots\ldots\ldots\ldots\ldots\ldots(8)$$

Where $z_t \sim N(0,1)$ and

$$h_t = \omega + \alpha \varepsilon^2_{t-1} + \beta h_{t-1} \quad \ldots\ldots\ldots\ldots\ldots\ldots(9)$$

All the parameters are positive and $(\alpha + \beta)$ measures the persistence of volatility. In general $(\alpha + \beta) < 1$ and has been observed to be very close to 1. The effect of any shock in volatility decays at a rate of $(1 - \alpha - \beta)$.

In case of GARCH (p, q) model equation 9 becomes

$$h_t = \omega + \alpha_1 \varepsilon^2_{t-1} + \alpha_2 \varepsilon^2_{t-2} + \cdots + \alpha_q \varepsilon^2_{t-q} + \beta_1 h_{t-1} + \beta_2 h_{t-2} + \cdots + \beta_p h_{t-p}$$

$$= \omega + \sum_{i=1}^{q} \alpha_i \varepsilon^2_{t-i} + \sum_{i=1}^{p} \beta_i h_{t-i} \quad (10)$$

Since our research framework is multivariate in nature, we have utilized multivariate extension of GARCH model in our research.

**3.3.2 EGARCH:** It is a variant of GARCH model. Formal EGARCH (1, 1) model can be characterized by:

$$\log(h_t) = \omega + \beta \log(h_{t-1}) + \alpha \left|\frac{\varepsilon_{t-1}}{\sqrt{h_{t-1}}}\right| + \gamma \frac{\varepsilon_{t-1}}{h_{t-1}} \qquad (11)$$

The parameter $\alpha$ measures the magnitude of volatility clustering. As the conditional variance is measured in logarithmic form, it allows the coefficients to have negative values. The parameter $\gamma$ captures the leverage effect.

**3.3.3 Unit Root Test:** As the time series must be stationary for GARCH, EGARCH modelling, Augmented Dickey Fuller (ADF) test is conducted to check the presence of unit roots. For a univariate time series, $y_t$ the ADF test basically applies regression to the following model:

$$\Delta y_t = \alpha + \beta_t + \gamma y_{t-1} + \delta_1 \Delta y_{t-1} + \delta_2 \Delta y_{t-2} + \cdots + \delta_{p-1} \Delta y_{t-p+1} + \varepsilon_t \qquad (12)$$

Where $\alpha$, $\beta_t$ and $p$ are constant, coefficient on a time trend and lag order of autoregressive process. It corresponds to random walk model if $\alpha = 0$ and $\beta = 0$ constraints are imposed. Whereas using the constraint $\beta = 0$ corresponds to modelling random walk with a drift. The Unit Root Test is then conducted under null hypothesis (H$_o$) $\gamma = 0$ against alternative hypothesis (H$_1$) $\gamma < 0$. The acceptance of null hypothesis implies nonstationary.

To quantitatively judge the performance of these models Mean Squared Error (MSE), Correlation Coefficient (R) and Theil Inequality (TI) measures are obtained. They are computed using the following set of equations.

$$\text{MSE} = \frac{1}{N} \sum_{i=1}^{N} \{Y_{\text{act}}(i) - Y_{\text{pred}}(i)\}^2 \quad \ldots\ldots\ldots\ldots\ldots\ldots\ldots\ldots\ldots\ldots\ldots\ldots (13)$$

$$R = \frac{\sum_{i=1}^{N}(Y_{\text{act}}(i) - \overline{Y_{\text{act}}})(Y_{\text{pred}}(i) - \overline{Y_{\text{pred}}})}{\sqrt{\sum_{i=1}^{N}(Y_{\text{act}}(i) - \overline{Y_{\text{act}}})^2} \sqrt{\sum_{i=1}^{N}(Y_{\text{pred}}(i) - \overline{Y_{\text{pred}}})^2}} \quad \ldots\ldots\ldots\ldots\ldots (14)$$

$$\text{TI} = \frac{\left[\frac{1}{N}\sum_{i=1}^{N}\left(Y_{\text{act}}(i) - Y_{\text{pred}}(i)\right)^2\right]^{1/2}}{\left[\frac{1}{N}\sum_{i=1}^{N}Y_{\text{act}}(i)^2\right]^{1/2} + \left[\frac{1}{N}\sum_{i=1}^{N}Y_{\text{pred}}(i)^2\right]^{1/2}} \quad \ldots\ldots\ldots\ldots\ldots\ldots (15)$$

Where $Y_{act}(i)$ and $Y_{pred}(i)$ are actual observed and predicted value of $i^{th}$ sample. N is the sample size. Whereas $\overline{Y_{act}}$ and $\overline{Y_{pred}}$ denote the average of actual and predicted values of N samples. Values MSE must be as low as possible for efficient prediction; ideally a value of zero signifies no error or perfect prediction. Both R and TI Values range between [0, 1]. They should be close to 1 for good prediction while 0 implies no prediction at all.

**4. Results of ANN Based Modeling**

As discussed we have utilized two different ANN models, MLFFNN, a traditional model, and NARX, a tailor made tool for time series modelling. The entire dataset from 1.1.2009 to 8.4.2016 has been suitably partitioned into training, validation and test dataset (70%, 15% & 15%) for predictive modelling exercise. Performance of the respective models are evaluated using Mean Squared Error (MSE) and Correlation Coefficient (R) measures.

**4.1 Results of MLFFNN modelling:** Only one hidden layer has been used while number of neurons in hidden layer has been varied at four levels (10, 20, 30 & 40 number of neurons). Levenberg-Marquardt (LM), Scaled Conjugate Gradient (SCG), Conjugate Gradient with Powell-Beale Restarts (CGPB), Fletcher-Powell Conjugate Gradient (FPCG), Polak-Ribiére Conjugate Gradient (PRCG) as five different variants of backpropagation algorithms have been adopted for learning purpose. So total twenty (no. of levels of neurons × no. of learning algorithms) numbers of experimental trials are conducted. Other specifications are mentioned in the following table.

**Table 3**
MLFFNN parameter settings

| Sl. No. | Parameter | Data/Technique Used |
|---|---|---|
| 1. | Number of input neuron(s) | Eight |
| 2. | Number of output neuron(s) | One |
| 3. | Transfer function(s) | Tan-sigmoid transfer function (tansig) in hidden layer & purelin in output layer. |
| 4. | Proportion of training, validation and test dataset | 70:15:15 |
| 5. | Error function(s) | Mean squared error (MSE) function |
| 6. | Type of Learning rule | Supervised learning rule |

The following figure depicts the strength of association between the target and output (predicted) exchange rate on training, validation, testing and entire dataset. Correlation coefficient values on respective data have also been mentioned.

One sample MLFFNN structure out of 20 trials is shown below.

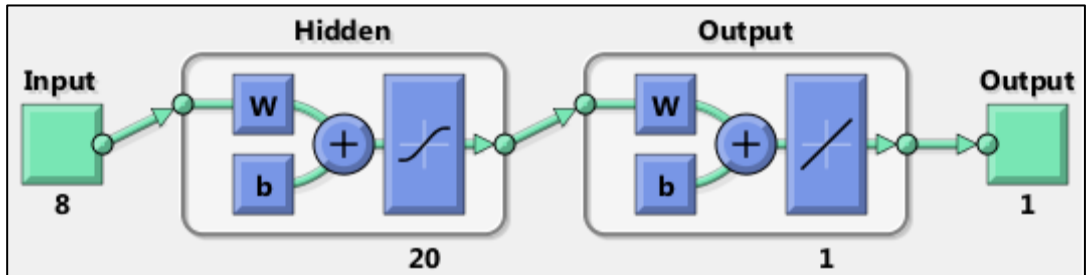

**Figure 3**
Sample MLFFNN Structure

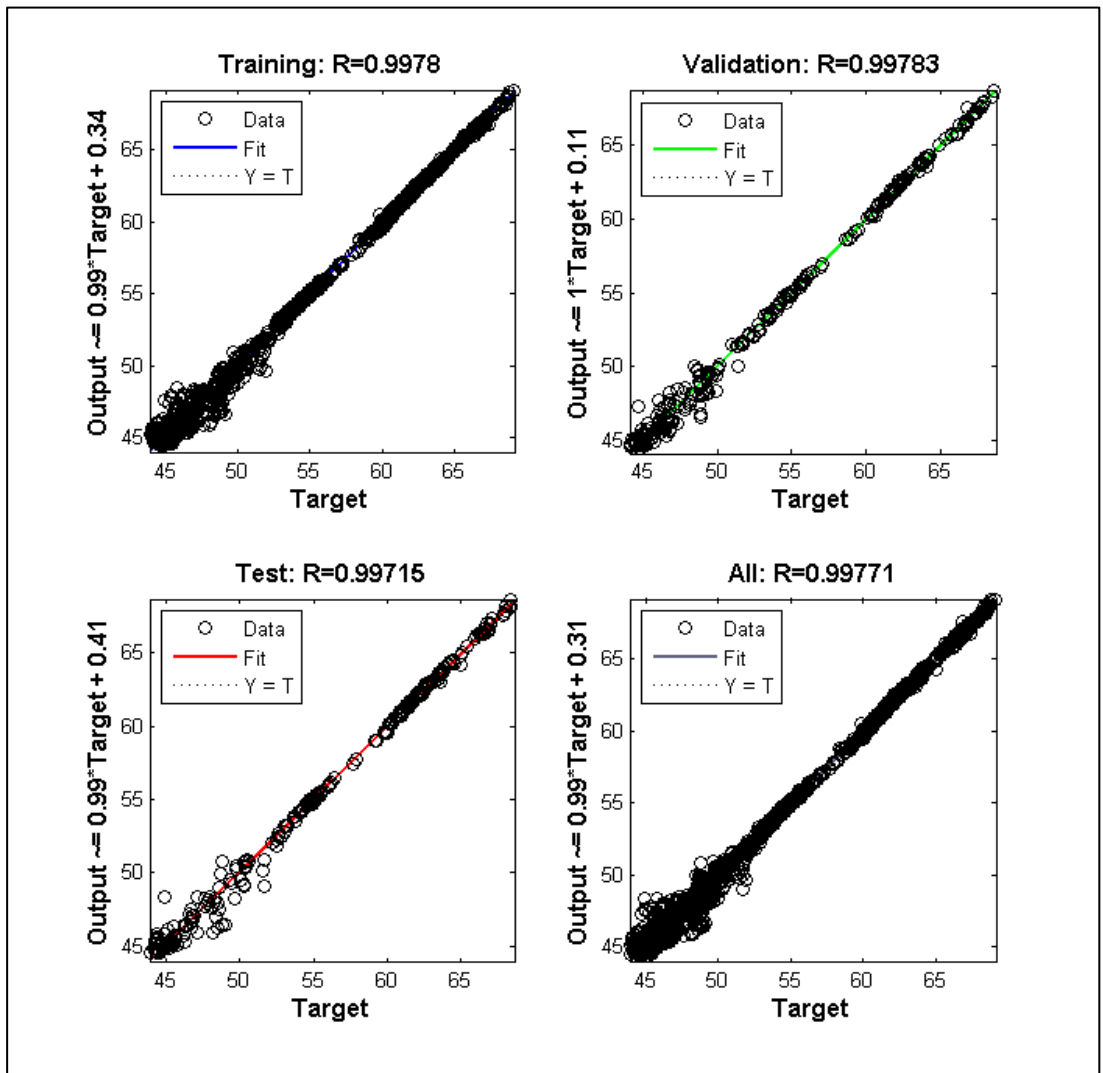

**Figure 4**
Performance of MLFFNN

It can be clearly seen that actual exchange rate (represented by target in the graph) and predicted exchange rate (represented by output in the graph) are very close for training, validation, test and entire data set. Almost a linear trend can be observed between the actual and predicted exchange rate which justifies the efficacy of the model. To further validate the claim MSE value is also obtained for individual experimental trials.

Statistics of MSE and R of predictive modelling on training and test dataset are summarized in following tables. For computation of MSE actual and predicted values have been rescaled to [0, 1] for all models.

**Table 4**
Performance on Training Dataset

| Statistics | R | MSE |
|---|---|---|
| Min | 0.9916 | 0.000132 |
| Max | 0.9989 | 0.000364 |
| Average | 0.9974 | 0.000227 |

**Table 5**
Performance on Test Dataset

| Statistics | R | MSE |
|---|---|---|
| Min | 0.9907 | 0.000169 |
| Max | 0.9981 | 0.000417 |
| Average | 0.9957 | 0.000289 |

High R values and negligible MSE values for both training and test data set imply that exchange rate can effectively be predicted using MLFFNN architecture using FX4, DJIAR, NIFTYR, DR, HSR, COP, CV and IV.

**4.2 Results of NARX modelling:** Similar to MLFFNN, only one hidden layer is employed in NARX network too. Delay of 2 units to consider the lagged values of both dependent and independent variables have been considered for model building. Number of neurons in hidden layer is varied at four levels and five learning algorithms have been used.

For the considered problem, general formulation of NARX structure as indicated in equation 6 is replaced by equation 16.

$FX(t) = f(FX4(t), FX4(t-1), FX4(t-2), NIFTY(t), NIFTY(t-1), NIFTY(t-2), DJIAR(t), DJIAR(t-1), DJIAR(t-2), HSR(t), HSR(t-1), HSR(t-2), DR(t), DR(t-1), DR(t-2), COP(t), COP(t-1), COP(t-2), CV(t), CV(t-1), CV(t-2), IV(t), IV(t-1), IV(t-2))$ ………………….. (16)

Similar to MLFFNN, five backpropagation algorithms namely, Levenberg-Marquardt (LM), Scaled Conjugate Gradient (SCG), Conjugate Gradient with Powell-Beale Restarts (CGB), Fletcher-Powell Conjugate Gradient (CGF), Polak-Ribiére Conjugate Gradient (CGP) are used for training. Other specifications of NARX are same as of MLFNN highlighted in table 3. One sample NARX structure is displayed in figure 5.

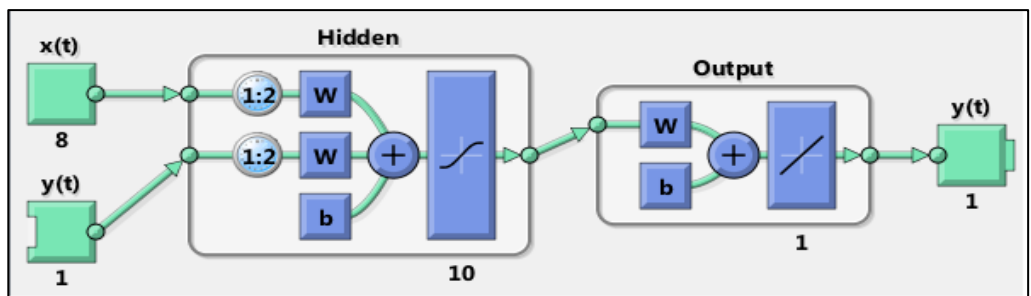

**Figure 5**
Sample NARX structure

Figure 6 graphically represents association of the actual and predicted exchange rate for all training, test and validation sample. Magnitude of error expressed as difference between actual and predicted values is also shown in the same figure. Although visual representation strongly suggests goodness of fit of NARX network in predicting exchange rate, to quantitatively justify the claim, MSE and R values are computed for training and test dataset.

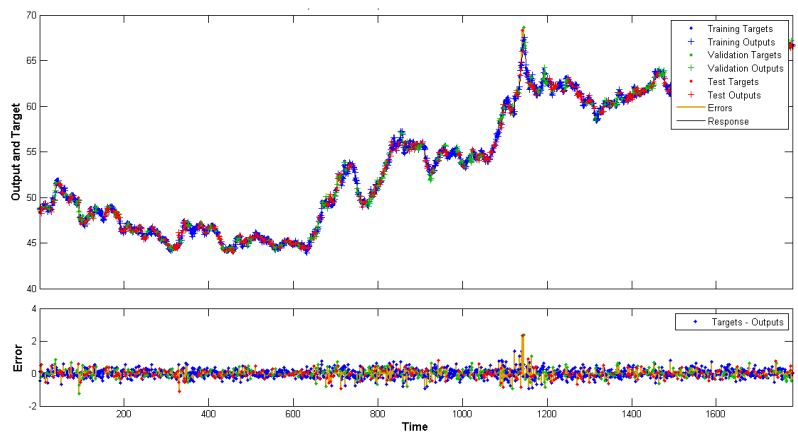

**Figure 6**
Visualization of NARX performance

Statistics of MSE and R of predictive modelling performance of NARX network on training and test dataset for all experimental trials are summarized in following tables.

**Table 6**
Performance on Training Dataset

| Statistics | R | MSE |
|---|---|---|
| Min | 0.9911 | 0.000125 |
| Max | 0.9979 | 0.000373 |
| Average | 0.9957 | 0.000229 |

**Table 7**
Performance on Test Dataset

| Statistics | R | MSE |
|---|---|---|
| Min | 0.9898 | 0.000169 |
| Max | 0.9942 | 0.000392 |
| Average | 0.9923 | 0.000271 |

It is evident from negligible MSE and high R values that the presented NARX network with 2 delay units has predicted exchange rate as a nonlinear function of FX4, DJIAR, NIFTYR, HSR, DR, COP, CV and IV quite effectively.

Error histogram with 20 bins is displayed below.

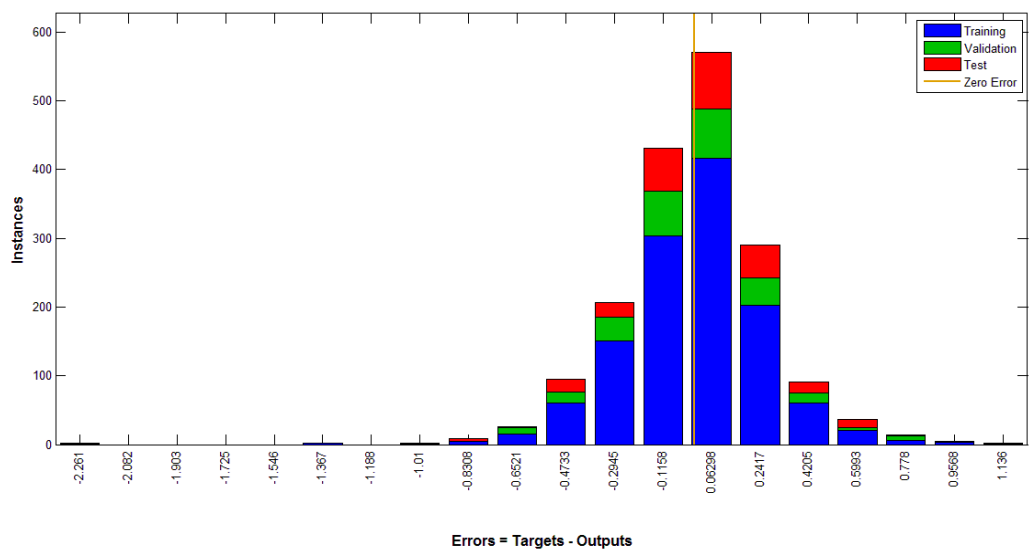

**Figure 7**
Error histogram of NARX modelling

**4.3 Assessment of Parameters**

To determine the impact of number of neurons in hidden layer and different backpropagation algorithms on performance in terms of MSE of MLFFNN and NARX on test data set, Analysis of Covariance (ANCOVA) has been performed. Different algorithms and number of neurons are treated as fixed factor and covariate respectively. Results are summarized in table 8.

**Table 8**
Tests of Between-Subjects Effects (MLFFNN)

Dependent Variable: MSE

| Source | Type III Sum of Squares | Df | Mean Square | F | Sig. | Partial Eta Squared |
|---|---|---|---|---|---|---|
| Corrected Model | 3.662E-008 | 5 | 7.324E-009 | 2.010 | .139 | .418 |
| Intercept | 1.644E-007 | 1 | 1.644E-007 | 45.130 | .000 | .763 |
| Neurons | 9.604E-011 | 1 | 9.604E-011 | .026 | .873 | .002 |
| Algorithms | 3.652E-008 | 4 | 9.131E-009 | 2.506 | .089 | .417 |
| Error | 5.101E-008 | 14 | 3.643E-009 | | | |
| Total | 1.118E-006 | 20 | | | | |
| Corrected Total | 8.763E-008 | 19 | | | | |

It is observed that varying the number of neurons in hidden layer does not have significant impact on predictive performance of MLFFNN. On the other hand, usage of different backpropagation algorithms has somewhat influence (at p-value < 0.1 level) on the performance. It can be inferred that number of neurons in hidden layer can be fixed at any of the four levels considered in this. The same proposition cannot be made for the various back propagation algorithms deployed though. In future, further investigations can be made using advanced Taguchi's experimental design methods or Response Surface Methodology to find the optimum level parameter settings.

**Table 10**
Tests of Between-Subjects Effects (NARX)

Dependent Variable: MSE

| Source | Type III Sum of Squares | Df | Mean Square | F | Sig. | Partial Eta Squared |
|---|---|---|---|---|---|---|
| Corrected Model | 5.731E-009 | 5 | 1.146E-009 | .202 | .956 | .067 |
| Intercept | 3.032E-007 | 1 | 3.032E-007 | 53.549 | .000 | .793 |
| Neurons | 3.745E-009 | 1 | 3.745E-009 | .661 | .430 | .045 |
| Algorithms | 1.986E-009 | 4 | 4.964E-010 | .088 | .985 | .024 |
| Error | 7.927E-008 | 14 | 5.662E-009 | | | |
| Total | 1.554E-006 | 20 | | | | |
| Corrected Total | 8.500E-008 | 19 | | | | |

For NARX model neither varying the number of neurons nor the usage of different training algorithms significantly affect the overall performance. Hence any specifications of parameters of NARX model out of twenty experimental setups can be suitably chosen for prediction of exchange rate.

## 5. Results of Time Series Based Modeling

As discussed, before proceeding with GARCH and EGARCH modelling to check whether the dataset is stationary or not, ADF test, discussed in section 3, is conducted. Additionally Philips-Perron (PP) test has been conducted too for the same. Similar to ADF test, acceptance of null hypothesis in PP test means the time series is nonstationary. Results are summarized in following table.

**Table 11**
Results of Unit Root Test (ADF)

| Variable | t-Statistic | p-value | Significant |
|---|---|---|---|
| **FX1** | -0.16374 | 0.9404 | Not Significant |
| **FX4** | -0.33843 | 0.9167 | Not Significant |
| **NIFTYR** | -39.82233 | 0.0000 | *** |
| **DJIAR** | -45.30706 | 0.0001 | *** |
| **HSR** | -42.06761 | 0.0000 | *** |
| **DR** | -41.16089 | 0.0000 | *** |
| **COP** | -42.82212 | 0.0000 | *** |
| **CV** | -4.221728 | 0.0006 | *** |
| **IV** | -3.676932 | 0.0005 | *** |

*** Significant at 1% level

**Table 12**
Results of Unit Root Test (Philips Perron Test)

| Variable | Adj. t-Statistic | p-value | Significant |
|---|---|---|---|
| **FX1** | -0.33662 | 0.917 | Not Significant |
| **FX4** | -0.43354 | 0.901 | Not Significant |
| **NIFTYR** | -39.7629 | 0.0000 | *** |
| **DJIAR** | -45.304 | 0.0001 | *** |
| **HSR** | -42.0675 | 0.0000 | *** |
| **DR** | -41.1526 | 0.0000 | *** |
| **COP** | -42.8303 | 0.0000 | *** |
| **CV** | -13.9375 | 0.0000 | *** |
| **IV** | -4.53662 | 0.0002 | *** |

*** Significant at 1% level

Both ADF and PP tests suggest that FX1 and FX4 are nonstationary. As only FX1 and FX4 are found to be nonstationary, we have taken the first order difference of these two variables and further applied ADF test to check the stationary constraint before building GARCH and EGARCH models.

**Table 13**
Results of Unit Root Test (First Difference Series) via ADF

| Variable | t-statistic | p-value | Significance |
|---|---|---|---|
| FX1 | -31.87125 | 0.0000 | *** |
| FX4 | -32.81953 | 0.0000 | *** |

*** Significant at 1% level

**Table 14**
Results of Unit Root Test (First Difference Series) via Philips Perron Test

| Variable | Adj. t-Statistic | p-value | Significance |
|---|---|---|---|
| FX1 | -40.5455 | 0.0000 | *** |
| FX4 | -41.0302 | 0.0000 | *** |

*** Significant at 1% level

From above two tables, FX1 and FX4 are identified as I(1). Subsequently ARCH Lagrange Multiplier (LM) test is performed. LM test statistic values and corresponding p-values for different duration of lags are reported in table 15.

**Table 15**
Results of ARCH LM Test

| Lag | F-statistic | p-value | Significance |
|---|---|---|---|
| 1-2 | 3077.471 | 0.0000 | *** |
| 1-5 | 1270.281 | 0.0000 | *** |
| 1-10 | 639.3180 | 0.0000 | *** |
| 1-15 | 425.0291 | 0.0000 | *** |

*** Significant at 1% level

As LM test statistic values are significant, presence the ARCH effect is deduced. These findings justifies the incorporation of GARCH and EGARCH in this research problem. We have utilized GARCH (1, 1), GARCH (2, 2), EGARCH (1, 1) and EGARCH (2, 2) models for forecasting purpose. Model fitness in terms of R-squared, Adjusted R-squared, Akaike Information Criterion (AIC), Schwarz Information Criterion (SIC) and Hannan-Quinn Information Criterion (HQC) are calculated for respective models and mentioned in table 16 and 17.

**Table 16**
Results of GARCH Model

| Model | R-squared | Adjusted R-squared | AIC | SC | HQC |
|---|---|---|---|---|---|
| **GARCH(1,1)** | 0.971979 | 0.971853 | -0.299478 | -0.262522 | -0.285829 |
| **GARCH(2,2)** | 0.973544 | 0.973424 | -0.292701 | -0.249585 | -0.276777 |

**Table 17**
Results of EGARCH Model

| Model | R-squared | Adjusted R-squared | AIC | SC | HQC |
|---|---|---|---|---|---|
| **EGARCH(1,1)** | 0.973361 | 0.973241 | -0.249598 | -0.209563 | -0.234812 |
| **EGARCH(2,2)** | 0.971791 | 0.971663 | -0.309553 | -0.263357 | -0.292491 |

As the values of the critical model indices are quite good, usage of volatility models for forecasting exchange rate is well justified. Subsequently estimated coefficient values of predictor variables by all four employed models are serially reported.

**Table 18**
Estimated Parameters of GARCH (1, 1) Model

| Variable | Coefficient | Standard Error | z-Statistic | p-Value | Significance |
|---|---|---|---|---|---|
| **Intercept/Constant** | 0.133896 | 0.015425 | 8.680527 | 0.00000 | *** |
| **FX4** | 0.979148 | 0.000192 | 5106.78 | 0.00000 | *** |
| **NIFTYR** | -0.26702 | 0.153306 | -1.74172 | 0.00000 | *** |
| **DJIAR** | 0.587357 | 0.186016 | 3.157555 | 0.00000 | *** |
| **HSR** | 0.004415 | 0.109882 | 0.040177 | 0.968 | Not Significant |
| **DR** | 0.170601 | 0.115334 | 1.479195 | 0.1391 | Not Significant |
| **COP** | -0.03279 | 0.09851 | -0.33282 | 0.7393 | Not Significant |
| **CV** | 0.003903 | 0.000257 | 15.15883 | 0.00000 | *** |
| **IV** | 0.00138 | 0.000315 | 4.377546 | 0.00000 | *** |

\*\*\* Significant at 1% level

Results reveal that out of eight predictors used in mean equation of GARCH model, FX4, NIFTYR, DJIAR, CV and IV are statistically significant.

**Table 19**
Estimated Parameters of GARCH (2, 2) Model

| Variable | Coefficient | Standard Error | z-Statistic | p-Value | Significance |
|---|---|---|---|---|---|
| **Intercept/Constant** | -0.13044 | 0.012862 | -10.1412 | 0.00000 | *** |
| **FX4** | 0.981626 | 0.000174 | 5650.204 | 0.00000 | *** |
| **NIFTYR** | 0.499303 | 0.138783 | 3.59772 | 0.0002 | *** |
| **DJIAR** | 0.220602 | 0.205063 | 1.075773 | 0.282 | Not Significant |
| **HSR** | -0.00017 | 0.108063 | -0.00156 | 0.9988 | Not Significant |
| **DR** | 0.149937 | 0.12472 | 1.202196 | 0.2293 | Not Significant |
| **COP** | -0.14303 | 0.083851 | -1.70572 | 0.0881 | * |
| **CV** | 0.004391 | 0.000244 | 18.0167 | 0.00000 | *** |
| **IV** | 0.00779 | 0.000246 | 31.68688 | 0.00000 | *** |

*** Significant at 1% level

* Significant at 10% level

For GARCH (2, 2) model FX4, NIFTYR, CV and IV are found to be highly significant. COP is significant at 10% level. Unlike GARCH (1, 1), DJIAR has been marked as not significant. HSR and DR are not significant as well.

**Table 20**
Estimated Parameters of EGARCH (1, 1) Model

| Variable | Coefficient | Standard Error | z-Statistic | p-Value | Significance |
|---|---|---|---|---|---|
| **Intercept/Constant** | -0.12546 | 0.009768 | -12.8437 | 0.00000 | *** |
| **FX4** | 0.981681 | 0.000118 | 8349.087 | 0.00000 | *** |
| **NIFTYR** | 0.43457 | 0.117823 | 3.688328 | 0.00000 | *** |
| **DJIAR** | 0.178091 | 0.168117 | 1.05933 | 0.2894 | Not Significant |
| **HSR** | 0.01102 | 0.113189 | 0.097361 | 0.9224 | Not Significant |
| **DR** | 0.103022 | 0.113425 | 0.908286 | 0.3637 | Not Significant |
| **COP** | 0.009301 | 0.071747 | 0.129633 | 0.8969 | Not Significant |
| **CV** | 0.003717 | 0.000219 | 16.97872 | 0.00000 | *** |
| **IV** | 0.007858 | 0.000267 | 29.39315 | 0.00000 | *** |

*** Significant at 1% level

In EGARCH (1, 1) model, significant predictors are turned out to be FX4, NIFTYR, CV and IV. Rest four predictors does not have significant impact on exchange rate.

**Table 21**
Estimated Parameters of EGARCH (2, 2) Model

| Variable | Coefficient | Standard Error | z-Statistic | p-Value | Significance |
|---|---|---|---|---|---|
| **Intercept/Constant** | 0.167494 | 0.014549 | 11.51244 | 0.00000 | *** |
| **FX4** | 0.978905 | 0.000173 | 5659.589 | 0.00000 | *** |
| **NIFTYR** | -0.25991 | 0.142216 | -1.82758 | 0.0676 | * |
| **DJIAR** | 0.60209 | 0.153425 | 3.924324 | 0.0001 | *** |
| **HSR** | 0.085213 | 0.091592 | 0.930357 | 0.3522 | Not Significant |
| **DR** | 0.282888 | 0.098929 | 2.859516 | 0.0032 | *** |
| **COP** | -0.02665 | 0.066709 | -0.39949 | 0.6895 | Not Significant |
| **CV** | 0.003608 | 0.000284 | 12.69635 | 0.00000 | *** |
| **IV** | 0.000697 | 0.000284 | 2.449421 | 0.0143 | ** |

*** Significant at 1% level, ** Significant at 5% level, * Significant at 1% level

Apart from HSR and COP, rest six independent variables have significant impact on movement of exchange rate according to EGARCH (2, 2) model.

To visualize the forecasting results obtained from GARCH (1, 1), GARCH (2, 2), EGARCH (1, 1) and EGARCH (2, 2) the following figures are presented.

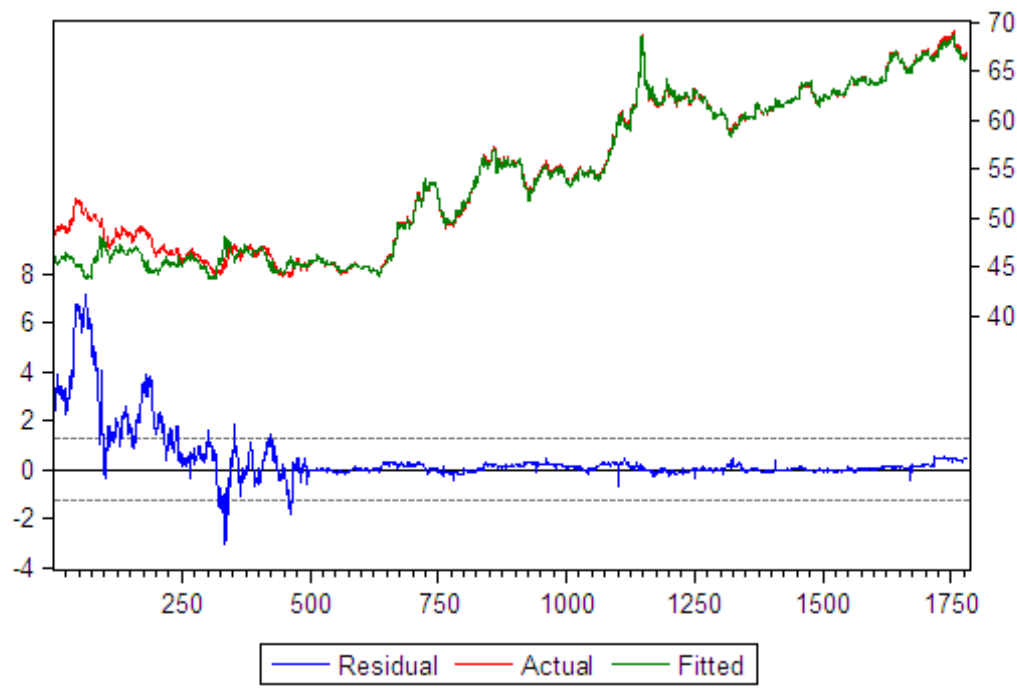

**Figure 8**
GARCH (1, 1) Performance

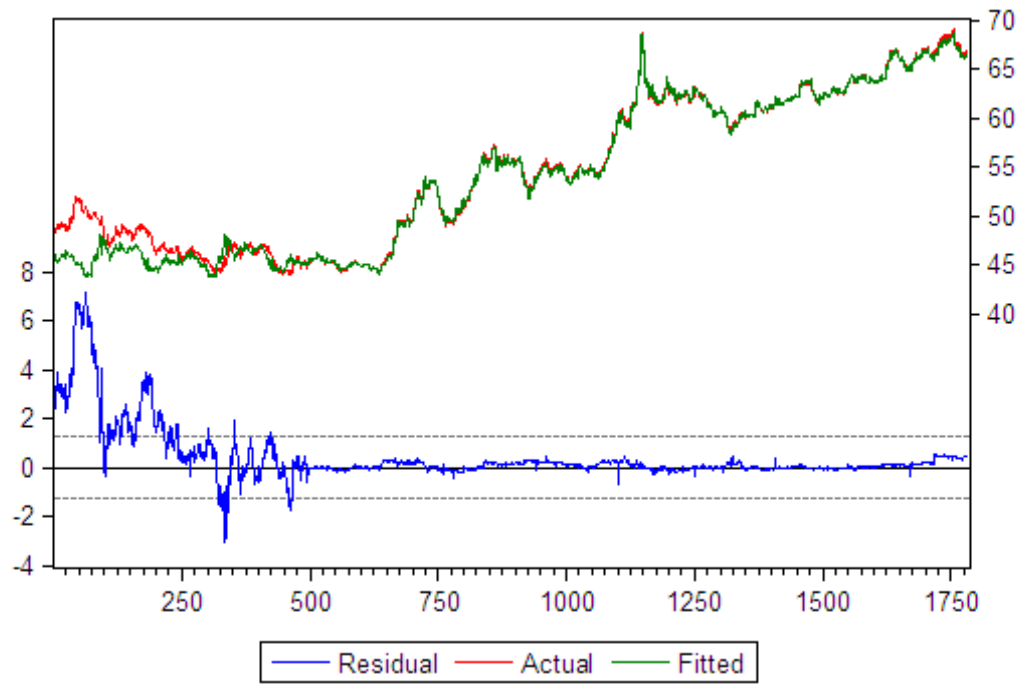

**Figure 9**
GARCH (2, 2) Performance

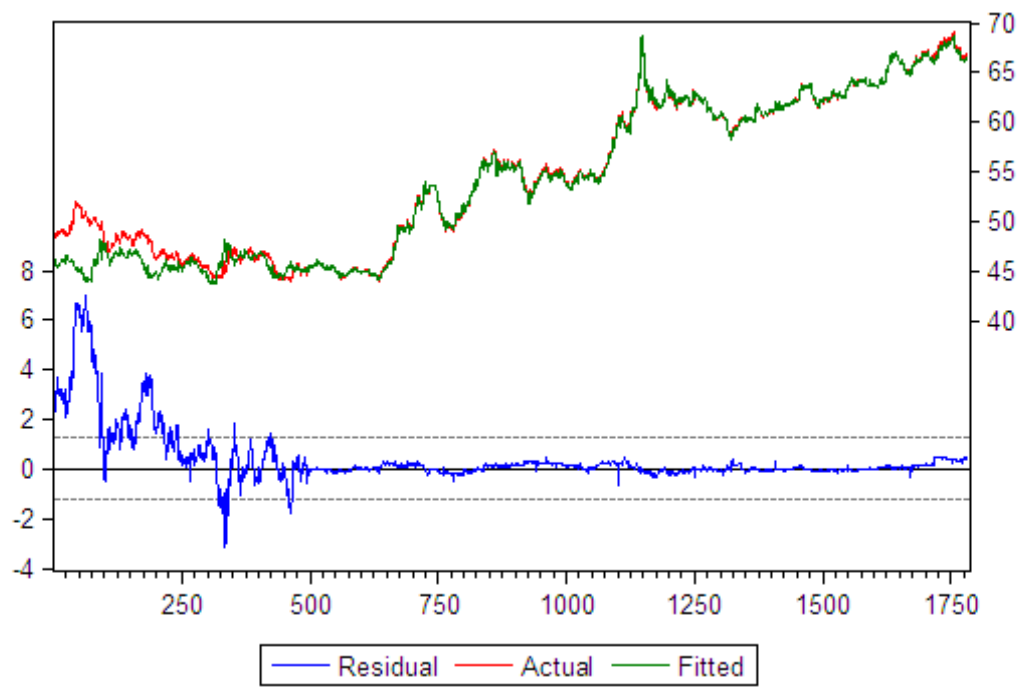

**Figure 10**
EGARCH (1, 1) Performance

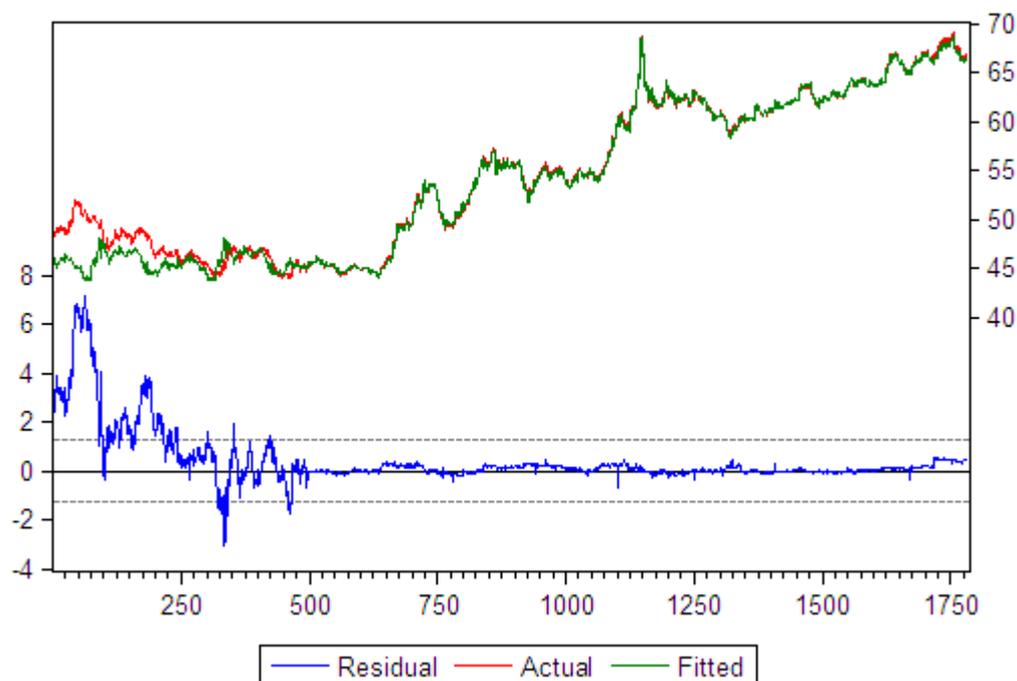

**Figure 11**
EGARCH (2, 2) Performance

Statistics of residuals are presented in Table 22. Conditional heteroscedasticity of residuals can be observed in terms of Kurtosis and Jarque-Bera Statistic that strongly justifies effectiveness of utilized time series framework.

**Table 22**
Residual Diagnostic

| Model | Mean | Median | Maximum | Minimum | Std. Dev. | Skewness | Kurtosis | Jarque-Bera | p-value |
|---|---|---|---|---|---|---|---|---|---|
| **GARCH (1, 1)** | 0.2809 | 0.7071 | 4.2983 | -5.4856 | 0.9600 | -0.6924 | 4.1962 | 248.4826 | 0.0000 |
| **GARCH (2, 2)** | 0.2779 | 0.7022 | 4.2727 | -5.1774 | 0.9610 | -0.7071 | 4.0186 | 225.4190 | 0.0000 |
| **EGARCH (1, 1)** | 0.2544 | 0.6309 | 4.5352 | -5.0639 | 0.9674 | -0.6969 | 4.1984 | 250.7210 | 0.0000 |
| **EGARCH (2, 2)** | 0.2788 | 0.6888 | 3.7100 | -5.4237 | 0.9597 | -0.6967 | 3.4505 | 159.1212 | 0.0000 |

Lastly for quantitative assessment of forecasting accuracy, MSE and Theil Inequality Coefficient are calculated using actual and obtained forecast values for all samples and shown in table 23.

**Table 23**
Forecasting Performance

| Model | MSE | Theil Inequality Coefficient |
|---|---|---|
| **GARCH (1, 1)** | 0.002310165 | 0.01152 |
| **GARCH (2, 2)** | 0.007260015 | 0.01550 |
| **EGARCH (1, 1)** | 0.00223638 | 0.01123 |
| **EGARCH (2, 2)** | 0.00235386 | 0.01552 |

Since both MSE and Theil Inequality Coefficient are substantially low, conclusions can be drawn that all the four models have been quite effective for forecasting exchange rate.

## 6. Comparative Analysis

In terms of MSE measures, performance of four GARCH family models and two ANN models are graphically plotted in figure 12.

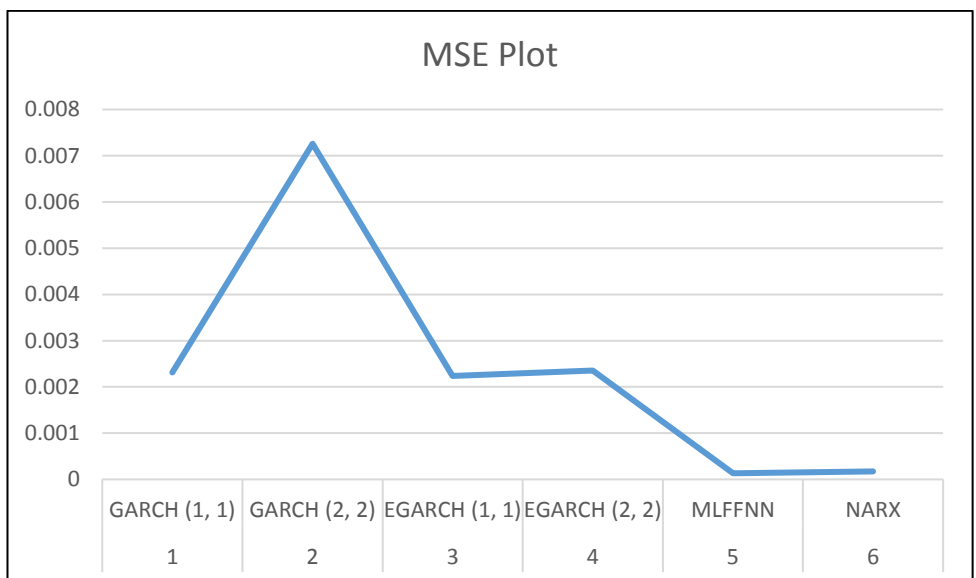

**Figure 12**
Comparative Analysis

Graphically it is quite evident that both the ANN models perform better than the four GARCH models. To statistically justify the claim t-test has been performed on MSE to determine whether the performance of ANN and GARCH family models are significantly different or not.

**Table 24**
t-Test Result (on test cases)

| p-Value | Significance |
|---|---|
| 0.0127 | ** |

** Significant at 5% level

As the t-Test statistic is significant at 5% level, it can be concluded that there is a significant difference between the performance of ANN and GARCH family models in predicting the exchange rates. ANN models are better than GARCH models in terms of MSE values.

## 7. Concluding Remarks

The paper uses both ANN based models and Econometric models in a multivariate framework to predict the Indian rupee US dollar exchange rate. The study is based on daily data. It incorporates explanatory variables from both the current account and the capital account of the balance of payments. During the process of generating results, it was observed that both sets of techniques generated useful and efficient predictions of the exchange rate. Further, the explanatory variables chosen were quite appropriate for the study. The application of both MLFFNN and NARX including the use of various backpropagation algorithms is quite unique and the non-linear relationship between the exchange rate and the explanatory variables have been effectively captured.

From the technique point of view, it is observed that the predictive performance of MLFFNN does not depend on the number of neurons in the hidden layer, but is sensitive to the backpropagation algorithms. For the NARX model, neither the number of neurons, nor the training algorithms, significantly affect the performance.

In econometric modelling, four different approaches namely, GARCH (1,1,), GARCH (2,2), EGARCH (1,1) and EGARCH (2,2) were used and the results have been reported. While the results obtained have been satisfactory, a comparative analysis of the ANN based models and the econometric models reveals that MLFFNN and NARX are better methods in terms of predictive efficiency.